\documentclass[aps,prl,twocolumn,showpacs]{revtex4-1}  % for review and submission
\usepackage{graphicx}  % needed for figures
\usepackage{dcolumn}   % needed for some tables
\usepackage{bm}        % for math
\usepackage{amssymb}   % for math
\usepackage{float}
\usepackage{color}
\usepackage{subfigure}

\begin{document}
\title {Role of Spin Flips versus Spin Transport in Non Thermal Electrons Excited by Ultrashort Optical Pulses in Transition Metals}

\widetext
\author {V. Shokeen}
\author {M. Sanchez Piaia}
\author {J.-Y. Bigot}
\email[E-mail: ]{bigot@unistra.fr}
\affiliation {Universit\'e de Strasbourg, CNRS, Institut de Physique et Chimie des Mat\'eriaux de Strasbourg, UMR 7504,  67034, Strasbourg, France}
\author{T. M\"uller}
\author{P. Elliott}
\author{J. K. Dewhurst}
\author{S. Sharma}
\email{sharma@mpi-halle.mp.de}
\author{E. K. U. Gross}
\affiliation{Max-Planck Institut f\"ur Microstrukture Physics, Weinberg 2, D-06120 Halle, Germany}

% \affiliation {Universit\'e de Strabourg, CNRS, Institut de Physique et Chimie des Mat\'eriaux de Strasbourg, (UMR 7504), % 67034, Strasbourg, France}

\date{\today}
%\begin{abstract}
% We show that the relaxation dynamics of non-thermal spins in ferromagnetic Ni and Co films induced by $10~fs$ optical
  %  pulses is different. While  spin flips due to spin-orbit interaction occur in both materials, the transport of
  %  super-diffusive electrons is different. Experimentally, the demagnetization signal, probed on both sides of
  % the films with absolute time reference, decreases monotonically in Ni up to $40~nm$ of spin propagation. In Cobalt a sign inversion of the demagnetization signal occurs between the front and back sides of the films, showing that
  % in Co majority spins has a larger diffusion component. This dissimilarity is explained theoretically by fully
  % ab-initio treatment of laser induced spin dynamics by means of Time Dependent Density Functional Theory.
  % Overall the primary instants of relaxation of non-equilibrium spins population involves  spin-orbit induced
  % spin flips, time dependent exchange interaction and super-diffusive transport, a mechanism which is highly
  % material dependent.
% \end{abstract}
\begin{abstract}
A joint theoretical and experimental investigation is performed to understand the underlying physics of laser-induced demagnetization in Ni and Co films. Experimentally dynamics of spins is studied by determining the time-dependent amplitude of the Voigt vector and theoretically {\it ab-initio} calculations are performed using time-dependent density functional theory. We demonstrate that overall spin-orbit induced spin-flips are the most significant contributors with super-diffusive spin transport playing very limited to no role. Our study highlights the material dependent nature of the demagnetization during the process of thermalization of non-equilibrium spins.
\end{abstract}

 \maketitle

In late 1990's it was shown that femtosecond optical pulses interacting with the magnetic matter leads to an ultrafast (time scale of $\sim100~fs$) macroscopic reduction in the magnetization \cite{Beaurepaire1996,Hohlfeld1997,Aeschlimann1997}. Several experiments have confirmed this finding and such a demagnetization has been broadly divided into two categories -- thermal demagnetization caused by hot electrons\cite{Beaurepaire1996,Hohlfeld1997,Aeschlimann1997,Koopmans2005,Schmidt2005,Radu2009,Rhi2003} and all-optical demagnetization and switching \cite{Kirilyuk2013} involving either non-compensated GdFeCo ferrimagnetic latices \cite{Hansteen2005}, the inverse Faraday effect in Garnets \cite{Kimel2005} or dichroic absorption in ferromagnetic multi-layers \cite{Lambert2014}. As the controllability of spins with light might strongly impact technological applications, with consequences for magnetic storage, spintronics, all-optical switching, heat assisted magnetic recording etc., this field of Femtomagnetism has recently become highly active, using diverse experimental approaches such as THz \cite{Kampfrath2011,Walowski2016}, Xray Circular Dichroism \cite{Stamm2007,Boeglin2010} or High Harmonic Generation \cite{La-O-Vorakiat2009}.

Despite this flurry of activity, underlying physics causing this ultrafast demagnetization still remains contested with some of the most prominent model used for explaining this demagnetization being-- the three temperatures model (3TM) \cite{Beaurepaire1996,Bigot2001}, Elliott-Yafet scattering induced spins-flips \cite{Koopmans2010}, non-thermal excitations \cite{Carva2013}, spin-orbit interaction induced spin-flips \cite{Zhang2000,Vonesch2012,Bigot2009,Krieger15,scirep16} and super-diffusive spin-transport \cite{Battiato2010}. This super-diffusive model relies on majority spin electrons diffusing away from (into the substrate) while minority spin electrons staying within the magnetic layers to cause a reduction in the moment. It is also very controversial as the results of the experiments by Vodungbo et. al \cite{Vodungbo2012} have been interpreted to confirm the assumptions of the model while the experimental data by Schelleken  et al. \cite{Schellekens2013} contest the validity of the very same assumptions.

In this article we present joint theory and experimental work in an attempt to resolve this controversy.
Experimentally systematic measurements of the ultrafast demagnetization and transport in Ni and Co thin films of different thicknesses using $10~fs$ optical pulses are performed. Magnetization dynamics is probed both at the front (where the laser pulse comes in) and at the back face of these magnetic films at various time delays.
Theoretically a full {\it ab-initio} study of laser induced spin dynamics in Ni and Co films (of various thickness) using time dependent density functional Theory (TDDFT)\cite{Runge1984} is performed. Both experiments and theory clearly suggest highly material dependent nature of the underlying physics of light induced demagnetization.
In Ni spin-flips dominate the physics of demagnetization at all times, while in Co the situation is more complex with spin-diffusion playing a significant role initially and spin-flips dominating the physics beyond first $\sim20~fs$.

{\it Experiment--} The experiments are performed with a modified pump-probe TRMO set-up (see Supplemental Material \cite{SupMat}) using $10~fs$ pump pulses focused onto the front face of the sample within a diameter of $40~\mu m$ and two $10~fs$ probe pulses focused on the front (F) and back (B) faces with a $30~\mu m$ diameter  and a density of energy ten times less than the pump. All beams are p-polarized with an accuracy of ${\pm}1^{\circ}$ with respect to the plane of incidence. The transmissions $T_{F,B}$, reflectivities $R_{F,B}$, Faraday rotations $\theta_{F,B}$, ellipticities $\eta_{F,B}$ and their corresponding time dependent differential quantities, with and without pump beam, $\frac{{\Delta}S_{F,B}}{S_{F,B}}(t)$ are measured as a function of the pump probe delay $t$. The temporal resolution is $0.5~fs$ using a grazing incidence mirror in the non-collinear pump-probes interferometer. Part of the reflected (F) and (B) probe beams are selected and interfere in a collinear Michelson interferometer to set the absolute arrival time of each pulses on the sample. The repetition rate of the laser is $80~MHz$, centered at $810~nm$ with a maximum density of energy  $0.5~nJ/pulse$ for the pump. All static or dynamical measurements are performed for the two opposite directions (${\phi}=0, 180^{\circ})$ of a static magnetic field of $3.5~k{\OE}$ perpendicular to the sample plane ${\phi}=90^{\circ}$. The direction of the initial unperturbed magnetization direction is obtained from the Stoner Wohlfarth model. The Ni and Co thin films with thicknesses varying between $10$ and $40~nm$ are grown by sputtering on a $500~{\mu}m-thick$ $Al_{2}O_{3}$ substrate and capped on the front face with $50~nm$ $Al_{2}O_{3}$.

The analysis of the experimental results requires to proceed in several steps, briefly summarized hereafter. The large spectral bandwidth of the pump and probe pulses requires first to retrieve the complex refractive index or equivalently the diagonal complex tensor $\widetilde{\epsilon}_{ii},~i=x,y,z$ from $R_{F,B}$ and $T_{F,B}$. The non-diagonal tensor elements $\widetilde{\epsilon}_{ij},~i{\neq}j$ are obtained from the boundary and propagation matrices in magneto-optical multi-layer films \cite{Zak1990} \cite{Robinson1964}, including the substrate and capping layers. A similar procedure is used for extracting the dynamical differential quantities. For the magnetization, the ultimate interesting quantity is the complex Voigt vector defined as: $\widetilde{Q}=-i{\widetilde{\epsilon}_{ij}/\widetilde{\epsilon}_{ii}}=Qe^{i\varphi_{q}}$. The modulus $Q$ is proportional to the magnetization. Naturally all dynamical quantities are obtained from the differential measurements with and without pump. $({\Delta}Q/Q)(t)$ obtained from the polar signals is therefore directly comparable with the calculated projection $S_{z}(t)$ of the magnetization along the direction 0z perpendicular to the Ni or Co samples planes 0xy  (${\phi}=90^{\circ}$).

As a typical representative set of measurements in a $10~nm-thick$ Ni sample, Figs. 1(a)-(c) show at short time delays (up to $150~fs$) the measured dynamical quantities $({\Delta}R_{F}/R_{F})(t)$, $({\Delta}T_{F}/T_{F})(t)$, $({\Delta}{\\\epsilon 1}_{xxF}/{\epsilon 1}_{xxF})(t)$, $({\Delta}{\epsilon 2}_{xxF}/{\epsilon 2}_{xxF})(t)$. $\epsilon 1_{xx}(t)$ and $\epsilon 2_{xx}(t)$ refer to the real and imaginary parts of the diagonal dielectric function. Similarly we extract the nondiagonal parts $\epsilon 1_{xy}(t)$ and $\epsilon 2_{xy}(t)$  allowing us obtaining the time dependent magnetization $({\Delta}Q_{F}/Q_{F})(t)$, $({\Delta}{\varphi_{q}}_{F}/{\varphi_{q}}_{F})(t)$. For this time scale up to $150~fs$ all differential quantities correspond to the thermalization dynamics of charges and spins. All curves are obtained for opposite magnetic fields and subtracted (respectively added), and divided by two, when they correspond to a quantity related to the non-diagonal (respectively diagonal) tensor. Near the delay $t=0$ the coherent spin-photon interaction is present \cite{Bigot2009}, clearly visible here because of the $10~fs$ ultra-short pump and probe pulses. Then the  magnetization $({\Delta}Q_{F}/Q_{F})(t)$  decreases to its minimum (Fig. 1(c)).  Figs. 1(d)-(f) show the same quantities up to $1.6~ps$ when the charges and spins relax to the lattice, leading to a partial re-magnetization (Fig. 1(f) left axis). These three curves are typical of the usual "thermal re-magnetization" that can be described by a 3TM or with spin-phonon scattering. In contrast the primary demagnetization induced by the $10~fs$ pulses clearly indicate that the spin-phonon interaction maybe discarded as already pointed out by Carva et al. \cite{Carva2013}.

\begin{figure}
  \centering
  \includegraphics[width=3.4 in,height=3 in, keepaspectratio]{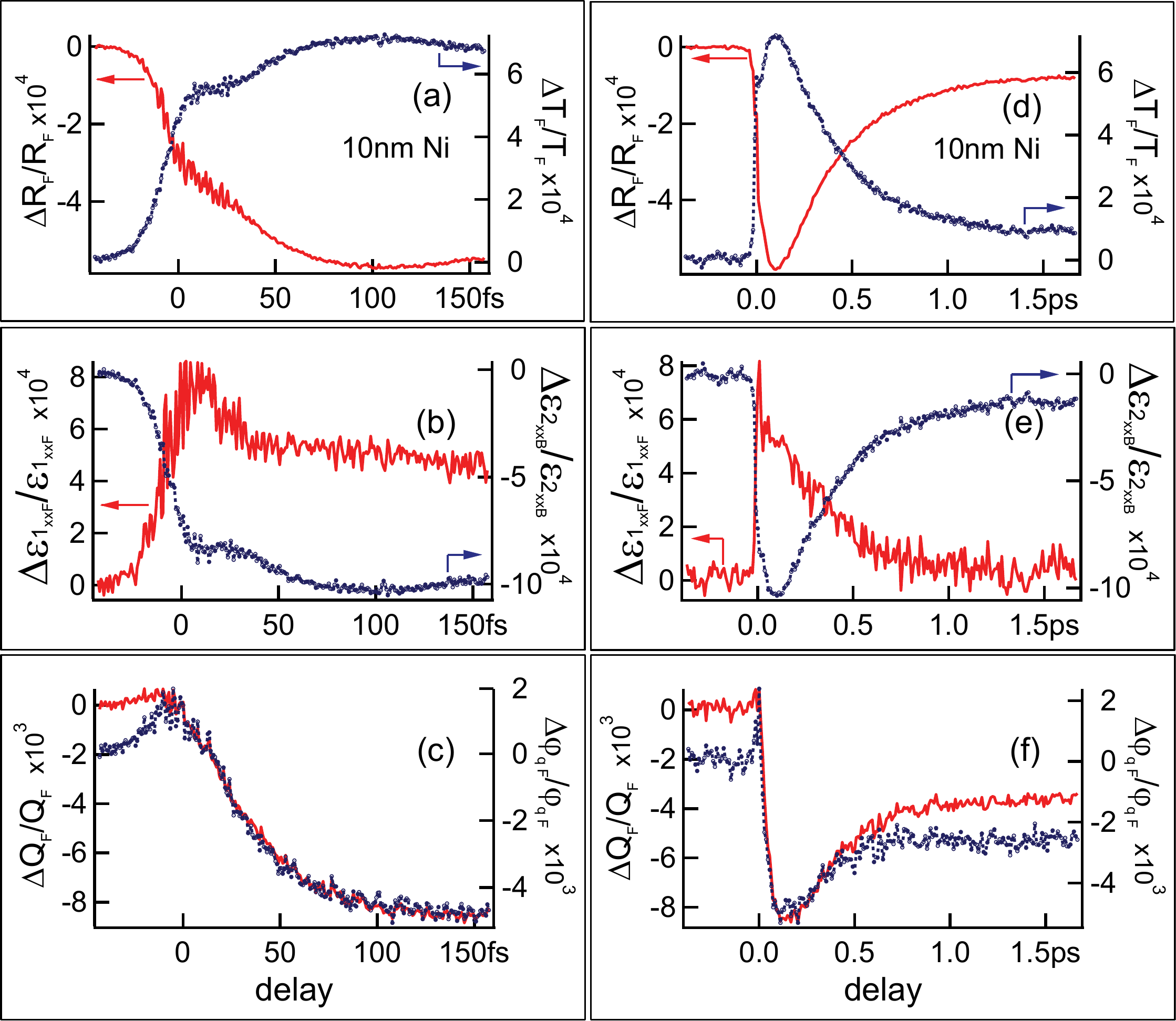}
 \caption{Ultrafast magnetization dynamics of $10~nm$ Ni film excited with $10~fs$ pump pulses. The sample is probed on the front face. Short delay dynamics of (a) : reflectivity $R_{F}$ (left axis) and transmission $T_{F}$ (right axis); (b):  Real and imaginary parts of diagonal dielectric tensor $\epsilon 1_{xxF}$ (left axis) and $\epsilon 2_{xxF}$ (right axis);  (c): Dynamics of amplitude $Q_{F}$ (right axis) and phase $\varphi_{F}$ (left axis) of the complex Voigt "vector". (d-f) shows the same quantities when the charges and the spins are relaxing to the lattice. The pump energy density is $5\times10^{-5}~Jcm^{-2}$.}
 \label{figure1}
\end{figure}

Let us now compare the effects of spin flips versus super-diffusive spin transport in Ni and Co samples. Towards that goal we have probed four samples, Ni and Co each with thicknesses $10$ and $40~nm$, both on the front and back sides (the pump pulse exciting always the front side). We focus only on the modulus of $Q$ $({\Delta}Q_{F}/Q_{F})(t)$ (left ordinate axis) and $({\Delta}Q_{B}/Q_{B})(t)$ (right ordinate axis) as they represent the magnetization dynamics. Figs. 2(a)-(b) show the results for the $10~nm$ and $40~nm$ Ni films at short delays. Fig. 2(c) shows the difference between the (B) and (F) faces. For the $10~nm$ film the demagnetization is larger on the back face ($\sim 1.9$ times), indicating that super-diffusive spins have propagated forward, but this propagation is in both spin channels and not just the majority spins as stipulated by the super-diffusive model. In contrast, for the $40~nm$ film (Fig. 2(b)), the demagnetization is less on the back side ($\sim 0.7$ times). This is better seen in Fig. 2(c) which clearly shows that for the $10~nm$ film the difference is negative while it is positive for the $40~nm$ film. Let us emphasize that all ${\Delta}Q/Q)(t)$ are negative quantities in this case (see Figs. 2(a) and 2(b)).

In the case of Cobalt the situation is very different. Figs. 2(d)-(e) show the results for the $10~nm$ and $40~nm$ Co films up to $300~fs$. A clear sign inversion occurs during the first $50~fs$ on the B face of the $10~nm$ film. This proves that a significant proportion of majority spins have propagated without spin flips. Instead, for the thicker $40~nm$ Co sample (Fig. 2(e)), the spin flips occur on both  (F) and (B) faces, showing that the majority spins are flipped after some propagation distance which we estimate to be $25\pm 3~nm$ by performing the same measurements on a $25~nm-thick$ sample (not shown here). This is also apparent in the differences of the (B) and (F) faces displayed in Fig. 2(f) (again remind that the quantity plotted is  $({\Delta}Q/Q)(t)$ for (B)-(F)). Thus, on the $10~nm$ Co film (Fig. 2(c)) one can see the contribution of the super-diffusive majority spins, which lead to the observation of a change in the sign of the magnetization in the early times.

\begin{figure}
  \centering
  \includegraphics[width=3.8 in, height=2.98 in, keepaspectratio]{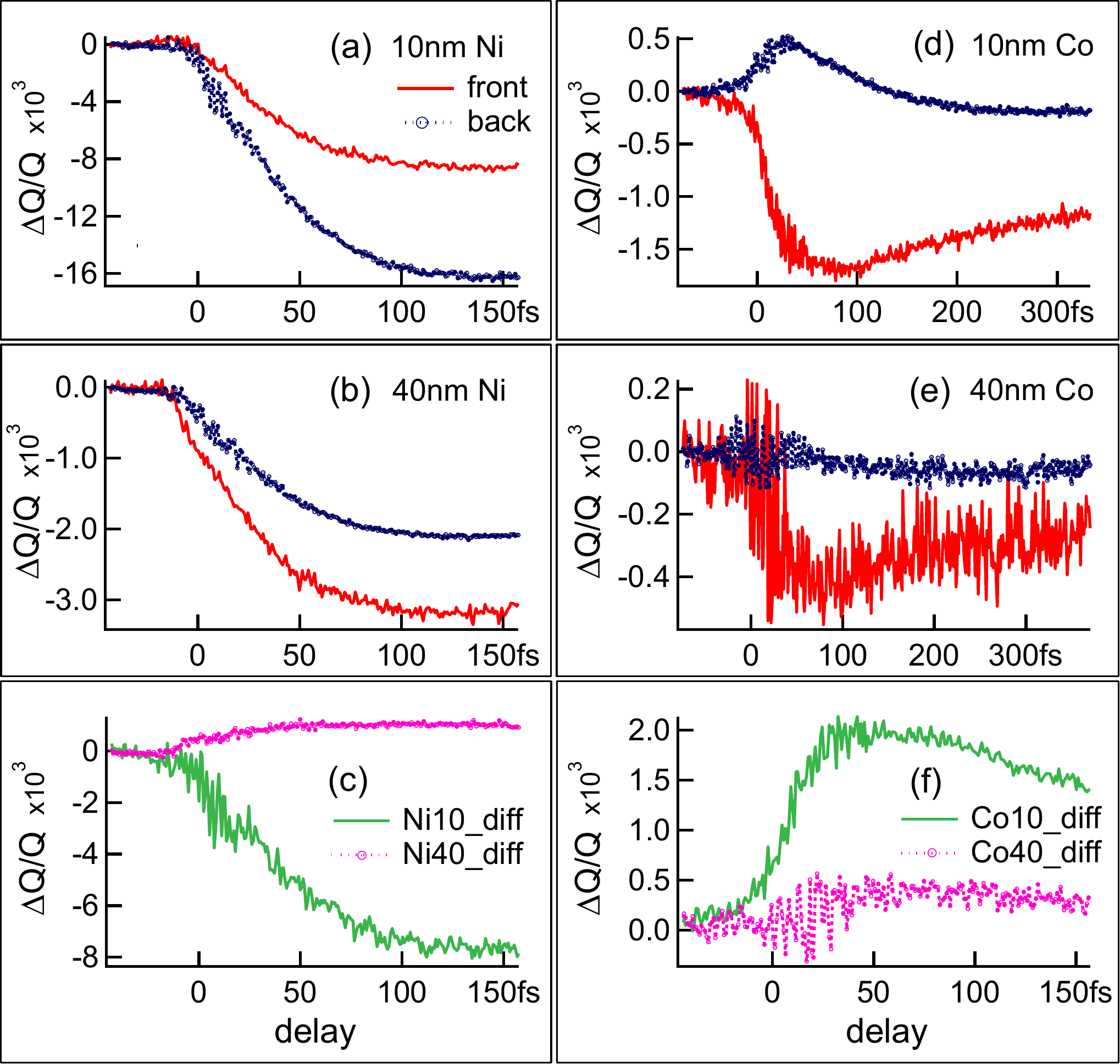}
  \caption{Ultrafast magnetization dynamics of Ni and Co films excited with $10~fs$ pulses probed on their front (F) and back (B) faces. Non-thermal regime in: (a) $10~nm$ Ni, (b) $40~nm$ Ni, (d) $10~nm$ Co, (e) $40~nm$ Co samples. (c) and (f) are the differences between (B) and (F) faces for Ni and Co samples respectively.}
 \label{figure2}
\end{figure}

\begin{figure}[ht]
\centerline{\begin{tabular}{cc}
\includegraphics[width=3.3 in, height=1.4 in, keepaspectratio]{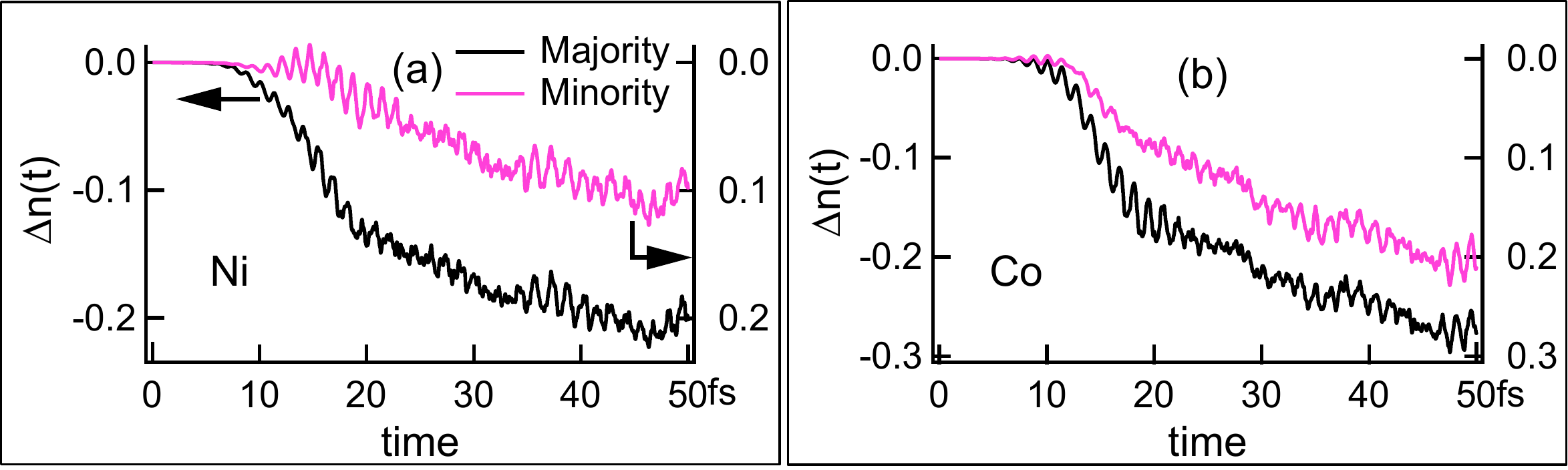}
\end{tabular}}
\caption{\label{updn}  Layer averaged majority (left axis) and minority (right axis) as a function of time (in fs).
Results for Ni are shown in the left panel and for Co in the right panel. As compared to the t=0 case there is decrease in majority and increase in minority spins.}
\end{figure}

{\it Theory--}  Super-diffusive model entails that the electrons in the majority spin channel are mobile and diffuse away from the magnetic layers while the electrons in the minority spin channel remain in the magnetic layers, leading to a local loss in the magnetic moment. This is equivalent to saying that the average majority charge in the magnetic layers shows a strong decrease (due to flow of majority spin-current) as a function of time while the averaged minority charge stays pretty much constant. Despite  totally neglecting the spin-orbit coupling, this model successfully explains experimentally observed demagnetization \cite{Battiato2010} in Ni. However, these demagnetization curves, being very simple, are easily reproducible using several other models as well \cite{Oppeneer2004,Zhukov2006} all of which rely on different underlying physics.

Given this what one requires is a fully {\it ab-initio} approach which does not make any assumptions about the underlying physics or the system under investigation. In the present work we have performed such first principles calculations using  TDDFT-- spin-orbit coupling is fully included, spins are treated in a fully non-collinear way and both the spin-channels  are treated at the same footing. This then allows for the effects of spin-current, spin diffusion, spin-flips due to spin-orbit coupling, restricted set of magnon excitations (by forming super-cell) and spin-canting (for details see Supplemental Material \cite{SupMat} and Ref. \cite{Krieger15,Dewhurst16}). In Fig. \ref{updn} are presented the results for the layer averaged change in the majority and minority charge as a function of time, $\Delta n(t) = n_{\rm maj/min}(t)-n_{\rm maj/min}(t=0)$ when all these processes are taken into account. From these results it is clear that change in minority spin electrons in the magnetic layers is almost as significant as majority and in total contrast with the super-diffusive model, both spin channels contribute strongly to the demagnetization process in magnetic films. This is in accordance with the experimental data of this work.

\begin{figure}[ht]
\centerline{\begin{tabular}{cc}
\includegraphics[width=3.25 in,height=3 in, keepaspectratio ]{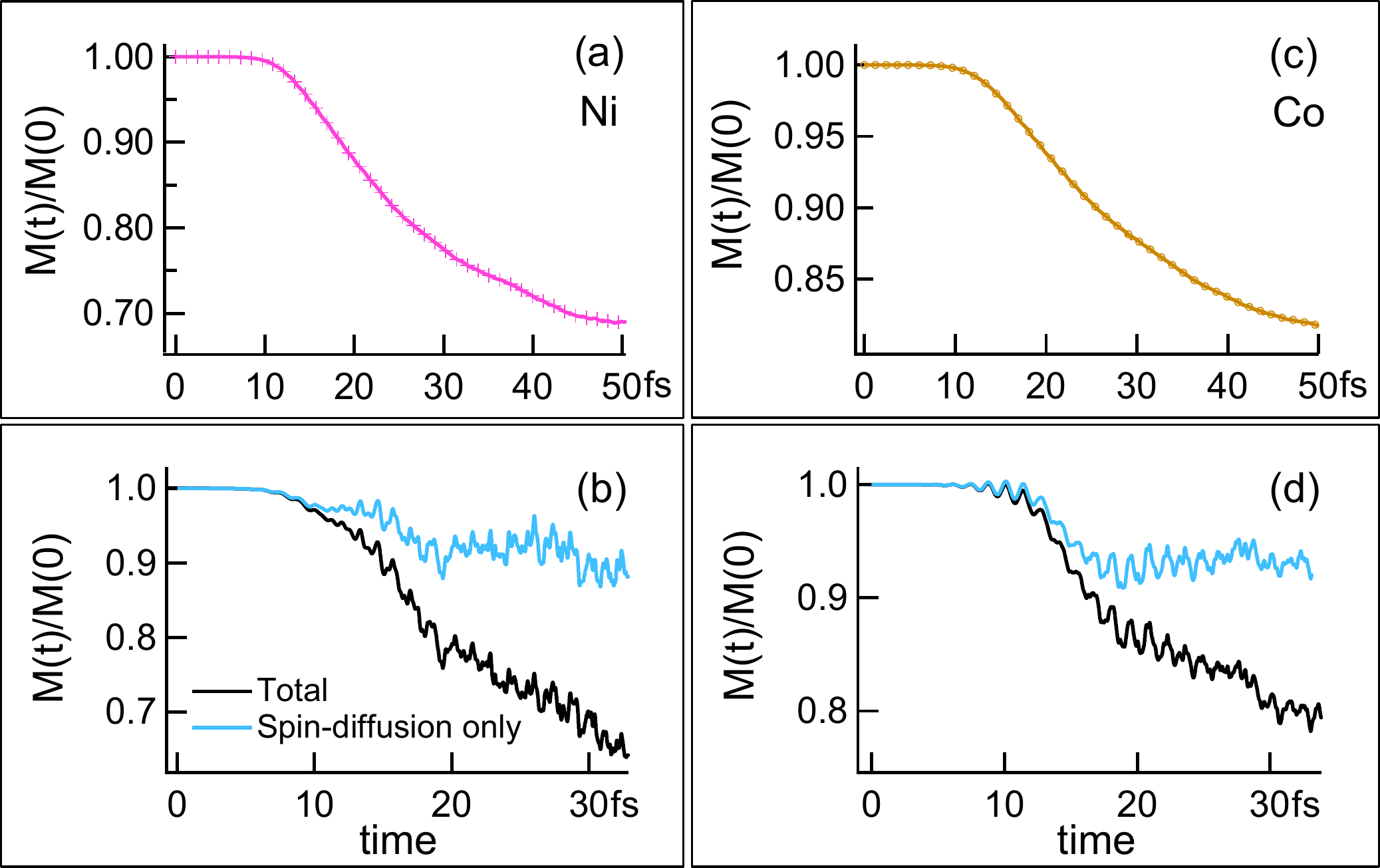}
\end{tabular}}
\caption{\label{noso} Total  (a) and (c) and layer resolved (b) and (d) normalized magnetic moment for Ni and Co films as a function of time (in fs). Layer resolved results are calculated in two ways-- (1)by time propagating the full Hamiltonian in Eq. (1) of ref. \cite{SupMat} (shown in black) and (2) by setting last term to zero in Eq. (1) of ref. \cite{SupMat} (shown in blue). The layer resolved data is for a representative layer (third layer) of a 7 mono-layer thick film.}
\end{figure}

In order to analyze these results in Fig. \ref{noso}, we present the total (Fig. \ref{noso} (a) and (c)) and layer resolved (Fig. \ref{noso} (b) and (d)) normalized moment for Ni and Co, $\frac{\rm M(t)}{\rm M(t=0)}$, as a function of time. The layer resolved results are obtained using two approaches: (1) by time propagating the full Hamiltonian in Eq. (1) of Ref. \cite{SupMat} and (2) by switching off the spin-orbit coupling term (setting last term to zero in Eq. (1) of Ref. \cite{SupMat}).  The later implies that demagnetization occurs only due to flow of spin-current from one part of the sample to another. This is similar to the scenario proposed by the super-diffusive model. While in the former case, together with the spin-current, spin-flips and spin-canting are also allowed. A comparison of the results from these two schemes would highlight the contribution of spin-diffusion alone to the total demagnetization.

From the top panels of Fig. \ref{noso} it is clear that, like experiments, we find Ni demagnetizes more than Co. The lower panels of this figure show that in the case of Ni the demagnetization caused by diffusion of spins alone strongly differs from the total demagnetization (which also includes the mechanism of spin-flips). This indicates that in the case of Ni spin-flips are the dominant mechanism for demagnetization. In the case of Co, in the early times (less than $20~fs$), a major part of demagnetization is caused by spin-currents (as proposed by the super-diffusive model). At times greater than $20~fs$, however, spin-flips start to become significant and, at larger times, ultimately dominate the physics of
demagnetization in Co as well.

This large temporal separation between start of spin-currents and spin-flips in Co ($\sim 20~fs$) could explain the experimental findings of this work-- in the early times a flow of spin-current causes an accumulation of majority spins at the back-face of Co films (leading to an increase in the moment) followed by which spin-flips become significant leading to a global demagnetization.
In total contrast to this for Ni the temporal separation between spin-currents and flips is small and spin-flips, which cause a global demagnetization, dominate the physics of demagnetization.  These results for Ni can explain not just the present experimental data but also the previous experimental work \cite{Schellekens2013}.

{\it Conclusions--} In conclusion we have performed a joint theory and experimental work to study thin films of Ni and Co excited and probed with $10~fs$ pulses.
Experimentally, time resolved magneto-optical study is performed and the magnetization dynamics is studied from the amplitude of the Voigt vector. Using samples of different thicknesses we study the significance of spin-flip vs the super-diffusive spin transport in the physics of demagnetization. Theoretically we employ state-of-the-art {\it ab-initio} method (i.e. time-dependent density functional theory) to study the magnetization dynamics of Ni and Co films.
From our work we conclude that (a) as opposed to super-diffusive spin transport, it is the spin-flips that play the most significant role in the process of demagnetization in both Ni and Co, (b) experimentally the front faces of both the materials both display a demagnetization behaviour as a function of time, (c) a sign inversion in the magnetization occurs at the back face of Co for early times ($t<50~fs$), while the back face of Ni shows same demagnetization behaviour as its front face and (d) this difference in the behaviour between the back faces of Co and Ni in early times can be explained based on our theoretical results which show a temporal separation between significant amount of spin-flips and majority spin-diffusion in Co. In Ni, on the other hand, both these processes occur at the same time. These results show that the demagnetization induced by femtosecond optical pulses in the two transition metals Ni and Co behave differently during the thermalization process of the spins.

{\it Acknowledgements-} JYB and VS acknowledge the financial support of the European Research Council, Advanced Grant  ATOMAG: (ERC-2009-AdG-20090325 247452) and the Agence Nationale de la Recherche in France, Equipex UNION (ANR-10-EQPX-52). SS and PE would like to thank DFG through SFB762 project and SS and TM would like to thank  QUTIF-SPP for funding.

\end{document}